\documentclass[showpacs,amssymb,floatfix]{revtex4}
\usepackage{graphicx}
\usepackage{dcolumn}
\usepackage{graphics}
\usepackage{epstopdf}
\usepackage{pdfpages}
\usepackage{bm}
\begin{document}
\title{All optical diode based on dipole modes of Kerr microcavity in asymmetric L-shaped photonic
crystal waveguide}
\author{E. N. Bulgakov and A. F. Sadreev$^*$}
\address{Kirensky Institute of Physics, Siberian Branch of
Russian Academy of Sciences, 660036, Krasnoyarsk, Russia $^*$Corresponding author: almas@tnp.krasn.ru}
\begin{abstract}
A design of all optical diode in $L$-shaped photonic crystal waveguide
is proposed that uses the
multistability of single nonlinear Kerr microcavity with two dipole modes.
Asymmetry of the waveguide is achieved by difference in coupling of the dipole modes
with the left and right legs of waveguide.
By use of coupled mode theory we present domains in axis of light frequency and amplitude where
an extremely high transmission contrast can be achieved. The direction of optical diode
transmission can be governed by power and frequency of injecting light.
The theory agrees with numerical solution of the Maxwell equations.
\end{abstract}
\maketitle

The optical bistability performance of a spatially asymmetric
photonic crystal (PhC) structure together puts the structure to be
a device that is called all-optical diode (AOD). Such a
device transmits light from one
side of the structure but not from other side. The optical diode has
an analogue with semiconductor diode that passes
electricity from one side only. AOD device
plays important role in all optical
signal processing. The first proposal for PhC-based optical
diodes was suggested by Scalora {\it et al} \cite{Scalora} based on the
dynamical shift of the photonic band edges in the 1D structure that consists
of alternate stacks of linear and nonlinear layers. After several years,
Gallo and Assanto \cite{Gallo} demonstrated that AOD can be
created by exploiting a $LiNiO_3$ waveguide with gratings.
Optical diode is characterized by the transmission
contrast $\eta =T_L/T_R$ where $T_L$ and $T_R$ are
the left and right  transmissions, respectively.

Up to now, many different mechanisms and methods to achieve optical
diodes have been proposed
(see Ref. \cite{Ding} and references there). Among of them nonlinear
cavities with different confinement
strengths along the left and right sides were used to give rise to
unidirectional transmission \cite{Mingaleev,Zhao,Lin}.
Because of the asymmetric confinement, the threshold for the transmission
through in-channel nonlinear cavities depends on the launch direction of the input
wave. However in spite of the merits of AOD structures, the transmission
contrast $\eta$ is usually low. In experiments by Gallo and Assanto the transmission contrast
was only $\eta=0.9$, although the CMT and FDTD simulations for single nonlinear cavity
showed $\eta=7.2$ \cite{Lin}.
Ultra high-contrast was demonstrated in  silicon Fano diode \cite{Ding}
$\eta=500$ with use of two uncoupled nonlinear micro cavities
in PhC directional waveguide with $T_L=0.7$ and $T_R=0.0014$.
In present paper we consider the AOD design exploring the
asymmetric PhC $L$-shaped waveguide with single
nonlinear microcavity at corner whose two degenerated dipole eigen modes lie in
the propagation band of the PhC waveguide. An asymmetry of the design is achieved
owe to different coupling strengths of dipole modes with propagating mode of
the waveguide in the left and right legs.
Such AOD  design demonstrates ultra high transmission contrast with $T_L=0.9$ and $T_R=0$
in the framework of CMT as well as
$T_L=0.95$ and $T_R=10^{-5}$  in the framework of numerical computation of the Maxwell equations
in PhC structure.

Unique symmetry properties
of dipole modes of linear cavity has been first
demonstrated in  cross waveguide in seminal paper by
Johnson {\it et al} \cite{johnson}.  Yanik {\it et
al} \cite{yanik} considered the nonlinear cavity of elliptic shape
defect rod with two dipole modes at the center of the cross photonic crystal
(PhC) waveguide. They have shown that due to a nonlinearity of the
cavity transmission over the x-direction can be reversibly
switched on/off by a control power over the y-direction to realize
all-optical transistor in X-shaped waveguide. Recently we have
shown that the nonlinear cavity with two degenerated dipole modes
positioned at the center of the directional waveguide can operate
in two regimes under transmission of even propagating mode \cite{BS}.
The first regime inherits the linear case when transmitting light excites only
the even dipole mode blocking outputs into perpendicular directions. However with
increasing of injecting power a bifurcation with excitement of the second odd dipole mode
occurs which emits light into perpendicular sections of the X-waveguide.
In the present letter we use this mechanism for optical diode.
The device exploits only single micro cavity what is important
with point of view of ultra compactness.

\begin{figure}
\includegraphics[height=8cm,width=8cm,clip=]{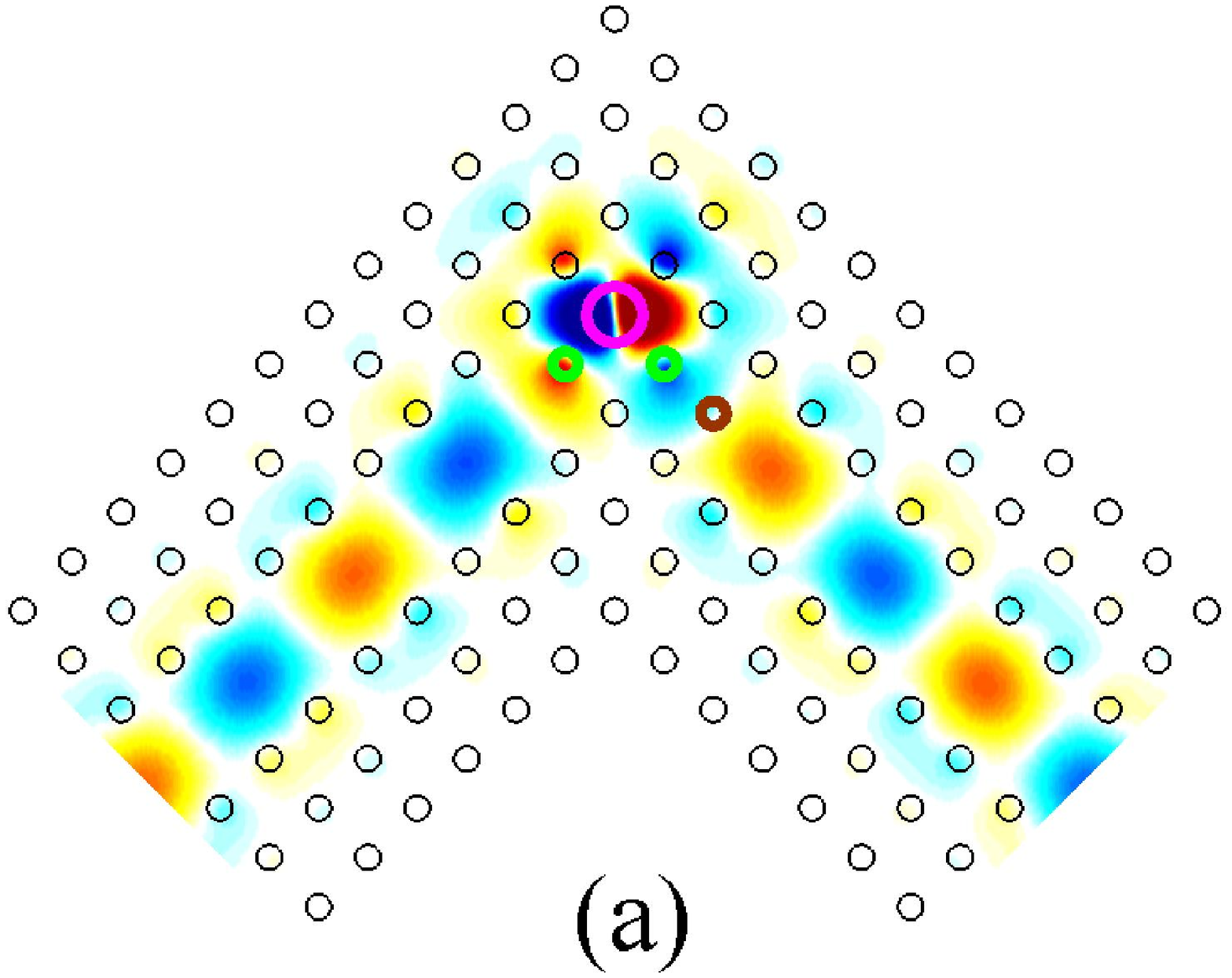}
\includegraphics[height=8cm,width=8cm,clip=]{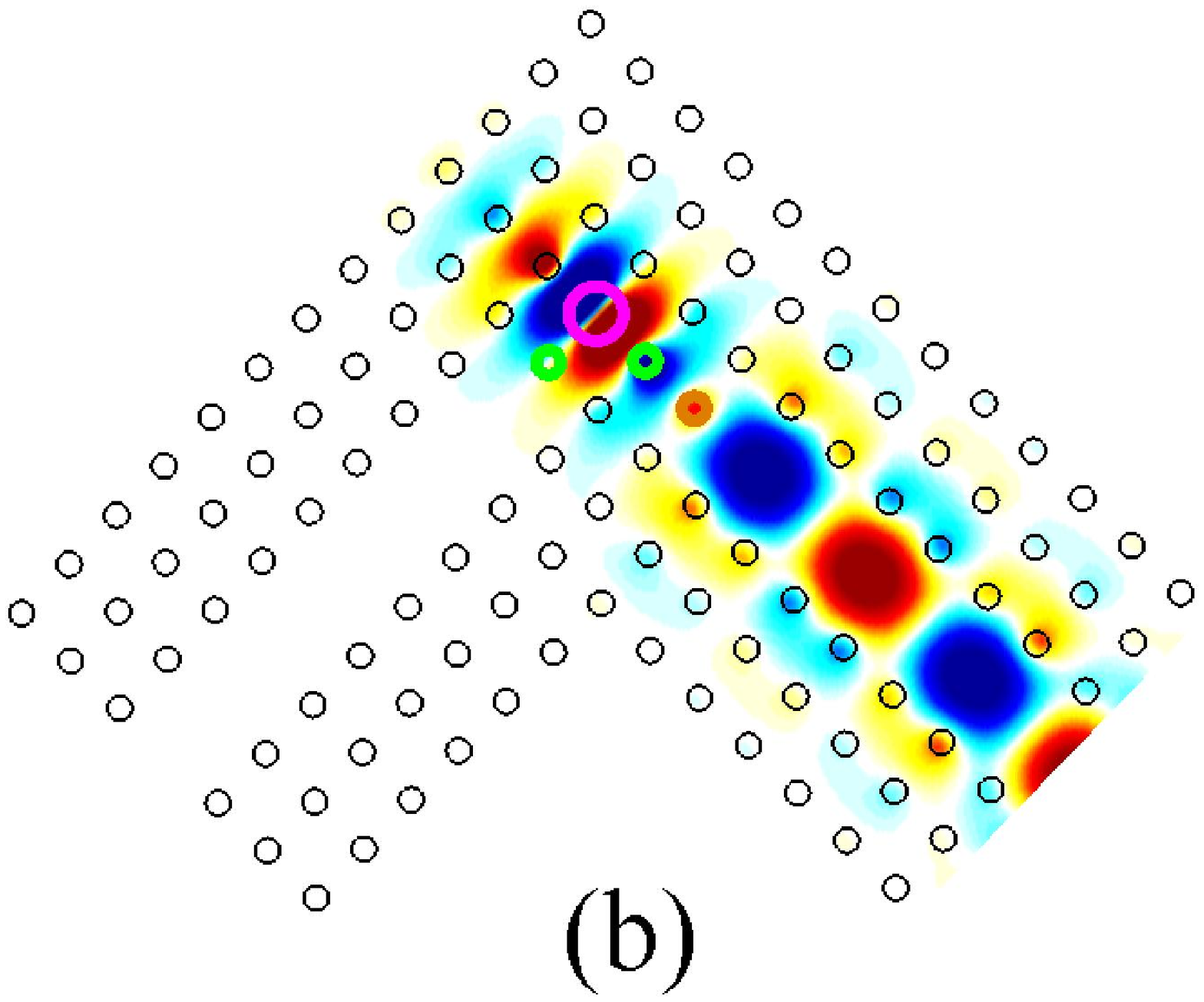}
\caption{(Color online) The nonlinear defect rod shown by
open pink larger circle has the radius $0.4a$ and $\epsilon_0=n_0^2=6.5$ and
the nonlinear refractive index $n_2=2\times 10^{-12}cm^2/W$
is placed into the corner of  the $L$ shaped waveguide formed by removal of
dielectric rods in the two-dimensional square lattice PhC. The PhC lattice  consisted of
the GaAs dielectric rods with radius $0.18a$ and dielectric
constant $\epsilon=11.56$ where $a=0.5\mu m$ is the lattice unit.
The additional two nearest rods (green in color) have radius $0.18a$ and $\epsilon=11.56$.
and third additional rod (brown in color) has radius $0.18a$ and $\epsilon=5$
is substituted into the right leg of the waveguide
in order to provide coupling asymmetry.} \label{fig1}
\end{figure}
We assume, the linear $L$-shaped PhC waveguide
is formed by removal of rows of dielectric rods as shown in Fig. \ref{fig1}. The
waveguide supports band of guided TM mode spanning from
the bottom band edge 0.315 to the upper one 0.41 in terms of $2\pi
c/a$ with the electric field directed along the rods.
This guided mode is even function along the line cross to waveguide.
The microcavity is formed by three linear rods and one nonlinear rod.
An asymmetric  design of the microcavity provides the asymmetric coupling with the left and right
legs of the waveguide. The material parameters of the cavity listed in Figure
caption are chosen
in so way that the dipole eigen frequencies of the microcavity belong the propagation band of the waveguide
while other eigen modes remain beyond.
The corresponding dipole modes
$E_1({\bf x})$ and $E_2({\bf x})$ obtained by numerical solution
of the Maxwell equations have ordinary shape presented, for example, in Refs. \cite{BS,joanbook}.
The dipole eigen frequencies in the design shown in Fig. \ref{fig1} equal
$\omega_1a/2\pi c=0.3658, \omega_2a/2\pi c=0.3650$.
For linear limit when injecting light is small such a dipole
microcavity will block a propagation of
the TM mode of light in the $L$ shaped waveguide ($T_L=T_R=0$) because of orthogonality of
the first/second dipole mode to the guided mode of right/left legs of the waveguide \cite{johnson}.

However in the case of the nonlinear dipole microcavity there might be a solution
when both dipole modes can be excited simultaneously because of
the nonlinear coupling between modes \cite{BS}. Therefore
for definite frequency and intensity of injected light such a system can
be opened for the light transmission. A window for the opening  mostly  depend
on a ratio between the nonlinearity constant and the coupling strengths of the dipole modes
of defect cavity  with propagating modes of the waveguide  \cite{BS}.
Because of the coupling strengths of the dipole modes with propagating mode are
different for the left and right
legs of the $L$ waveguide the thresholds for light transmissions
from the left to the right and back will be different \cite{Ding}. That is a key principle for the AOD in present PhC
nonlinear structure.

The coupled mode equations \cite{suh} for the nonlinear dipole mode amplitudes
have the following form \cite{BS}
\begin{eqnarray}\label{perturb}
i\dot{A_1}=[\omega_1+V_{11}-i\gamma_1/2]A_1+V_{12}A_2+
i\sqrt{\gamma_1}E_Le^{-i\omega t},\nonumber\\
i\dot{A_2}=[\omega_2+V_{22}-i\gamma_2/2]A_2+V_{21}A_1
+i\sqrt{\gamma_2}E_Re^{-i\omega t},
\end{eqnarray}
where $E_L$ and $E_R$ are the amplitudes of injecting light ingoing in
the left and right legs respectively, $A_1$ and $A_2$ are the amplitudes of
dipole modes, $\gamma_1, \gamma_2$ are the coupling strengths of the
first and second dipole modes with propagating modes of the left/right legs of the
L-shaped waveguide,
\begin{equation}\label{Vmn}
\langle m|V|n\rangle=-\frac{\omega_0}{2N_m}\int
d^2\vec{r}\delta\epsilon(\vec{r})E_m(\vec{r}) E_n(\vec{r}),
\end{equation}
\begin{equation}\label{deltaeps}
\delta\epsilon(\vec{r})=\frac{n_0cn_2|E(\vec{r})|^2}{4\pi}\approx
\frac{n_0cn_2|A_1E_1(\vec{r})+A_2E_2(\vec{r})|^2}{4\pi},
\end{equation}
is the nonlinear contribution to the dielectric constant of the
defect rod with instantaneous Kerr nonlinearity,
\begin{eqnarray}\label{nor}
 N_m=\int d^2\vec{r} \epsilon(\vec{r})E_m^2(\vec{r})=\frac{a^2}{cn_2}
\end{eqnarray}
is the normalization constant of the eigen-modes with
$\epsilon(\vec{r})$ as the dielectric constant of whole defectless
PhC.
After substitution of Eqs. (\ref{Vmn}) and (\ref{nor}) and
$A_m(t)=A_me^{-i\omega t}$ into the CMT equations
(\ref{perturb}) we write the stationary CMT equations in the dimensionless form
\begin{eqnarray}\label{A1A2dip}
&[\omega-\omega_1+\lambda_{11}I_1+\lambda_{12}
I_2+i\gamma_1/2]A_1+2\lambda_{12}
Re(A_1^{*}A_2)A_2=i\sqrt{\gamma_1}E_L,&\nonumber\\
&2\lambda_{12}Re(A_1A_2^{*})A_1+[\omega-\omega_2+\lambda_{22}I_2+\lambda_{12}
I_1+i\gamma_2/2]A_2=i\sqrt{\gamma_2}E_R,&
\end{eqnarray}
where we introduced $I_m=|A_m|^2$ as the intensities of the dipole
modes and dimensionless constants of nonlinearity
\begin{eqnarray}\label{lambda}
\lambda_{mn}=\frac{\omega_0n_0c^2n_2^2}{8\pi a^2}\int_{\sigma}
E_m^2(x,y)E_n^2(x,y)d^2\vec{r}
\end{eqnarray}
where $\sigma$ is the
cross-section of the defect rod. Respectively, the outgoing transmission
amplitudes to the left leg and to the right leg of the waveguide equal \cite{suh}
\begin{eqnarray}\label{TLR}
&t_L=\sqrt{\gamma_1}A_1-E_L&\nonumber\\
&t_R=\sqrt{\gamma_2}A_2-E_R.&
\end{eqnarray}

\begin{figure}
\includegraphics[height=4.2cm,width=4cm,clip=]{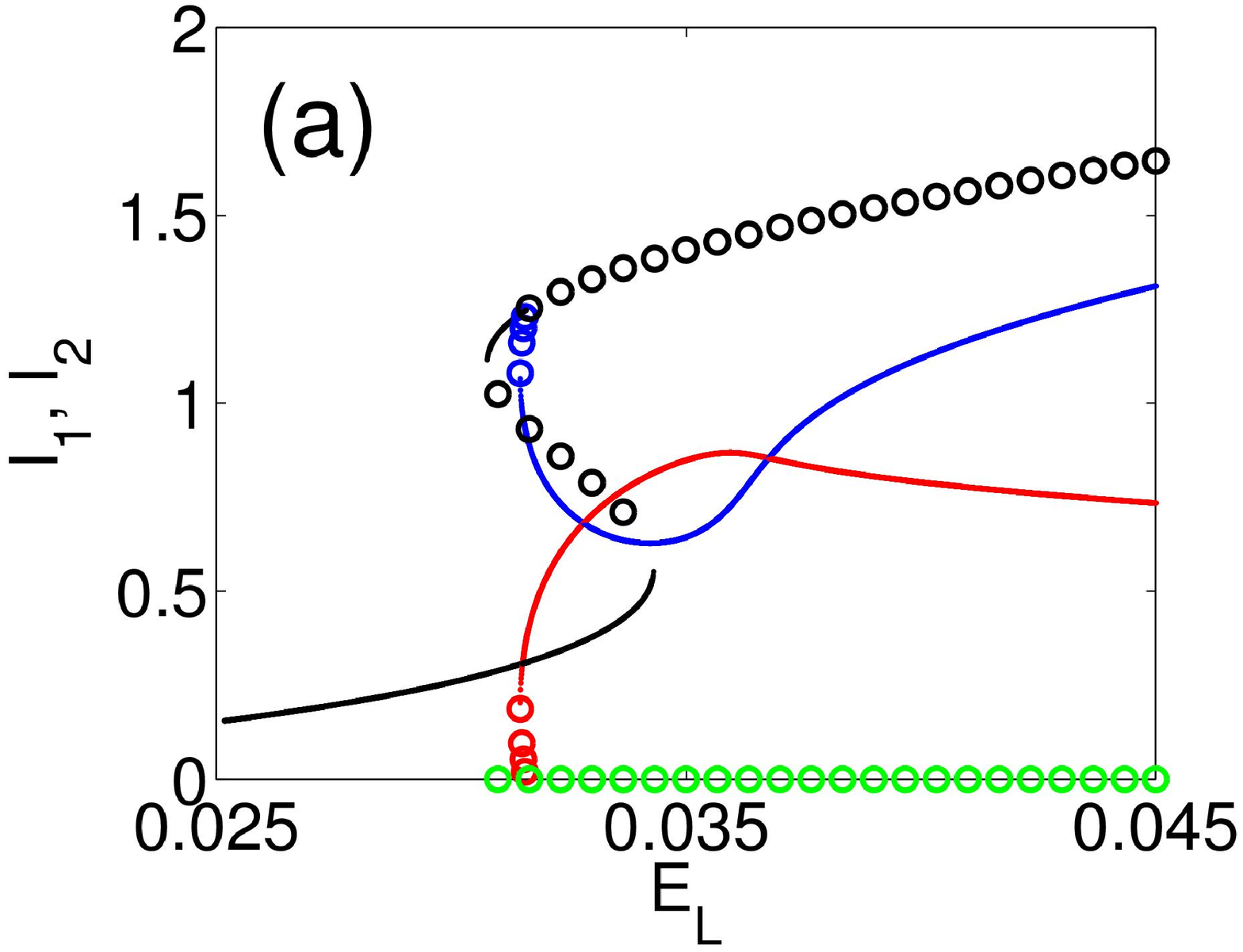}
\includegraphics[height=4cm,width=4cm,clip=]{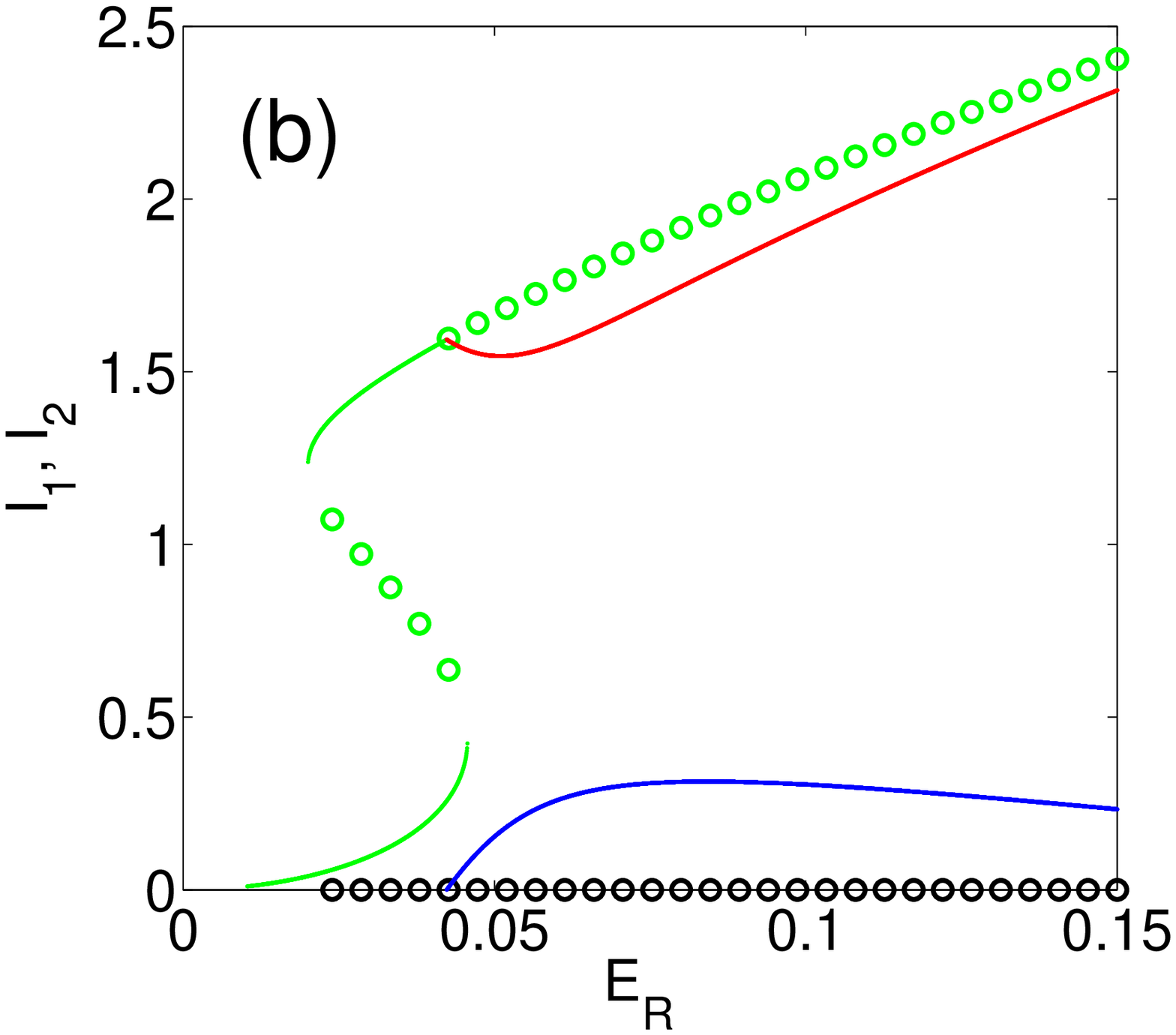}
\includegraphics[height=4cm,width=4cm,clip=]{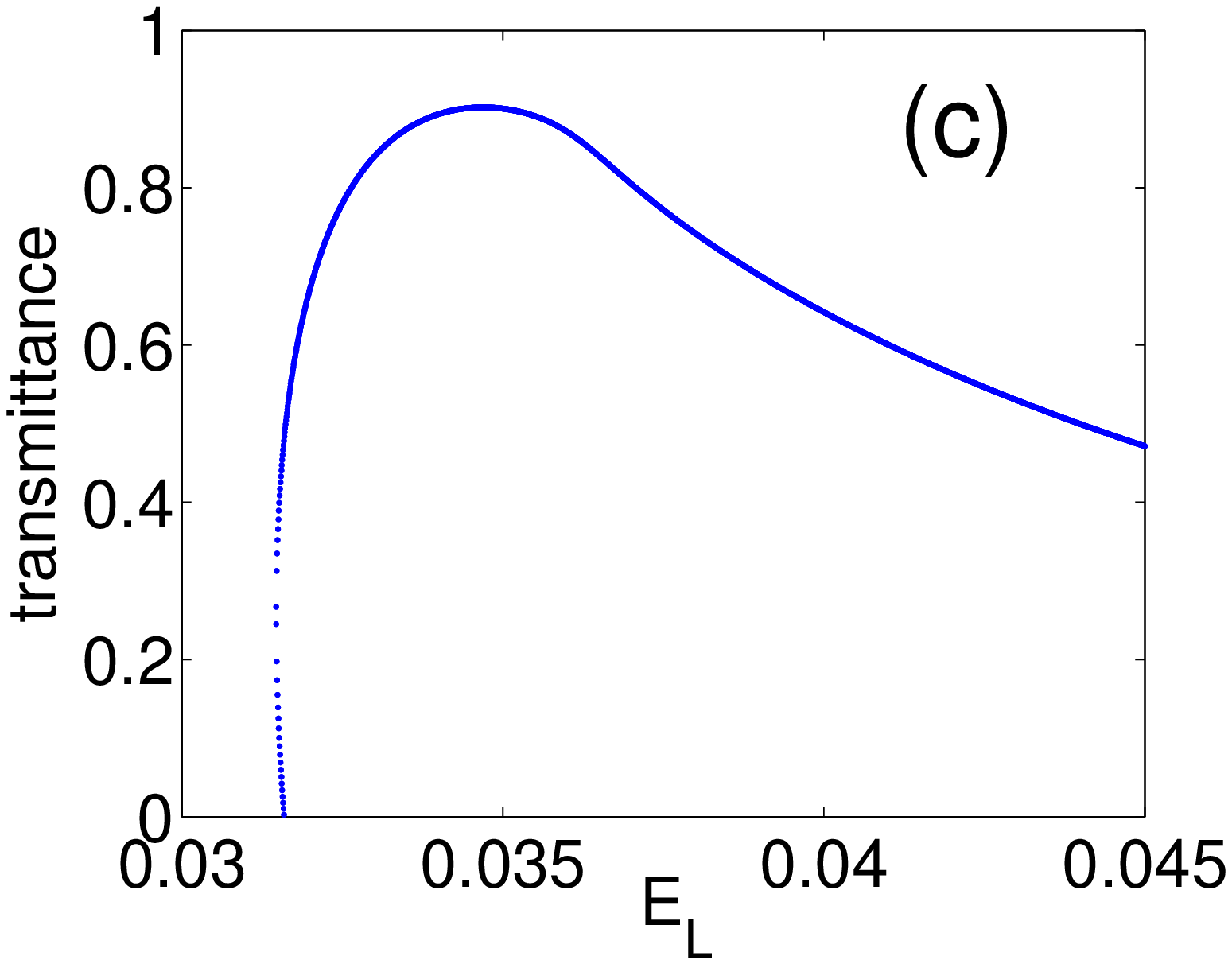}
\includegraphics[height=4cm,width=4cm,clip=]{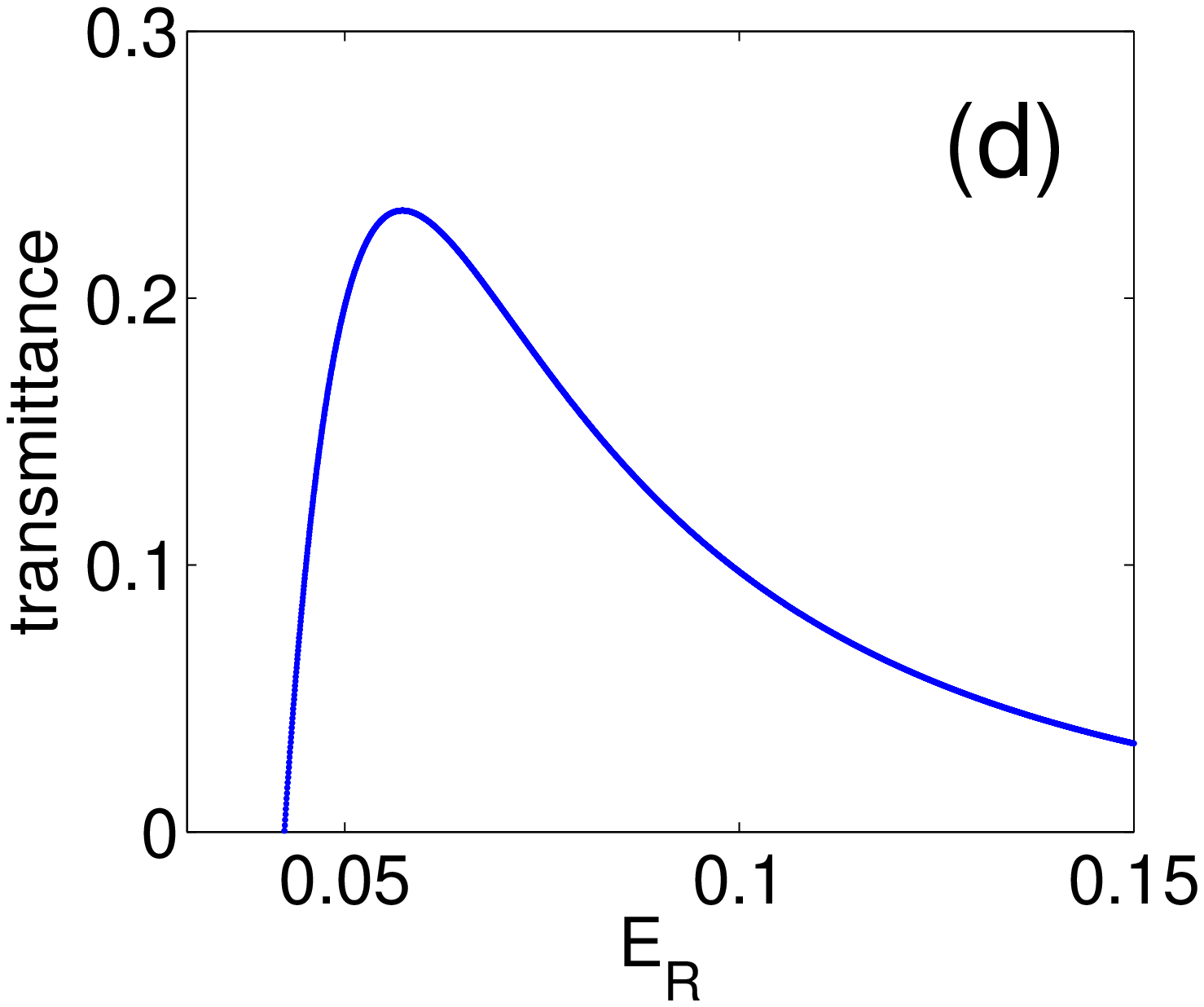}
\caption{(Color online) Intensities of dipole mode excitations vs light amplitude
injected from the left $E_L$ (a) and right $E_R$ (b) for $\omega=0.3618$.
Open circles mark unstable solutions of Eq. (\ref{A1A2dip}).
The ordinary solution inherited from the linear case which does not permit transmission
is colored by black ($I_1$) and green ($I_2$).
The bifurcated new solution intrinsic for nonlinear case opens transmissions colored
by blue ($I_1$) and red ($I_2$).
Transmittances corresponded to this bifurcated solution are shown in (c) and (d).}
\label{fig2}
\end{figure}

\begin{figure}
\includegraphics[height=6cm,width=7cm,clip=]{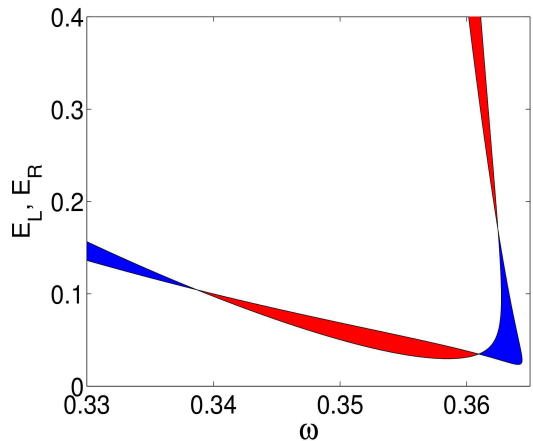}
\caption{(Color online)
The domains of AOD. The domains of AOD from the left to the right
is colored by blue while the domains of AOD from the right to the left is colored by red.}
\label{fig3}
\end{figure}

The results of numerical computation of the nonlinear CMT equations (\ref{A1A2dip})
altogether with Eqs. (\ref{TLR}) are presented in Fig. \ref{fig2} where we have
taken for simplicity the dipole modes are degenerated:
$\omega_1=\omega_2=0.3655$.
Substituting numerically calculated
eigen dipole modes into Eqs.(\ref{lambda})
we obtain $\lambda_{11}=\lambda_{22}=0.002963,
\lambda_{12}=0.001035$. We calculated $\gamma_1, \gamma_2$ via
widths of phase flips in reflection
amplitude \cite{Lee} for the linear dipole cavity.
Fig. \ref{fig2} (a) shows that for small amplitude injected from the left
the second dipole mode is not excited. However
when the amplitude $E_L$ exceeds the threshold value $E_{Lc}$ bifurcated solution arises with
both dipole modes
excited. As a result the transmission from the
left to the right is opened  while the transmission
from the right to the left is still closed as shown in Fig. \ref{fig2} (c, d).
Similar threshold exists for transmission from the right to the left when $E_R>E_{Rc}$.
However because of asymmetry of the design the threshold values for the right/left
injected lights $E_{Rc}$ and $E_{Lc}$ are different.
Thus, we obtained the AOD device
with infinite transmission contrast between $E_{Lc}$ and $E_{Rc}$.
From Eqs. (\ref{A1A2dip}) we can find the threshold values by use of
simple algebra from the condition that $I_2$ (Fig. \ref{fig2} (a)) or $I_1$ (Fig. \ref{fig2} (b))
tends to zero
\begin{eqnarray}\label{Ec}
    &\gamma_1|E_{Lc}|^2=I_{1c}[(\omega-\omega_0+\lambda_{11}I_{1c})^2+\gamma_1^2/4]&\nonumber\\
&\gamma_2|E_{Rc}|^2=I_{2c}[(\omega-\omega_0+\lambda_{11}I_{2c})^2+\gamma_2^2/4],&
\end{eqnarray}
where
\begin{eqnarray}\label{Ic}
&I_{1c}=\frac{2(\omega_0-\omega) \pm \sqrt{(\omega_0-\omega)^2-3\gamma_2^2/4}}
{3\lambda_{12}},&\nonumber\\
&I_{2c}=\frac{2(\omega_0-\omega) \pm \sqrt{(\omega_0-\omega)^2-3\gamma_1^2/4}}{3\lambda_{12}}.&
\end{eqnarray}
Answer to the question which direction
will be opened the first depends on sophisticated interplay between the resonance
widths, on the frequency and the nonlinearity constants.
Domains of AOD with infinite high transmission contrast which follow from
Eqs. (\ref{Ec}) and (\ref{Ic}) are shown in Fig. \ref{fig3} by red and green fields.

\begin{figure}
\includegraphics[height=6cm,width=8cm,clip=]{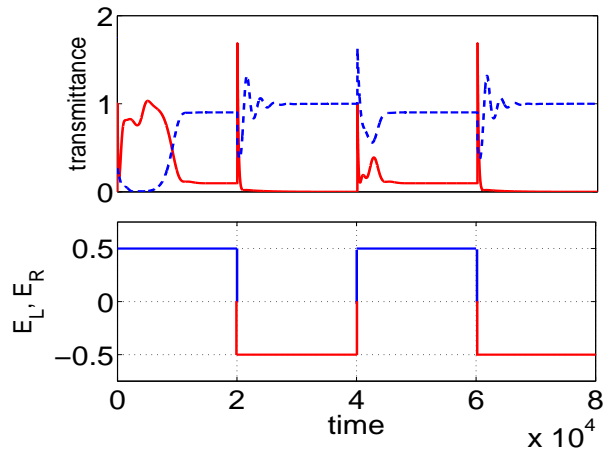}
\caption{(Color online) Time dependence of outputs to right leg (solid red line) and to
left leg (dash blue line) which follow the alternating injected impulses.
Below impulse launched from the left is marked by blue color and
impulse launched from the right is marked by red color.}
\label{fig4}
\end{figure}

What is important the stable bifurcated solution belongs to the domain of AOD
while the linear case inherited solution which blocks light transmission
might be unstable in this domain  as one can see from
Fig.  \ref{fig2} (a, b), in particular for $\omega=0.3618, E_L=0.35, E_R=0.35$.
That can be used for opening and closing of light transmission, i.e., for optical
switching. In Fig. \ref{fig4} we show
solution of temporal CMT equations (\ref{perturb}).
For the first impulse of light injected from the left with duration $2\times 10^4$ (in terms $a/2\pi c$)
 we observe at first weak transmission to the right (shown by dash blue line) which is substituted
by transmission close to 90\%. The next impulse injected from the right blocks the
output to the left for duration of impulse. The next series of impulses gives the same result.
Thus, two impulses of light
alternatingly injected from two legs of the L-shaped waveguide results in
AOD effect with infinite transmission contrast.

\begin{figure}
\includegraphics[height=5cm,width=7cm,clip=]{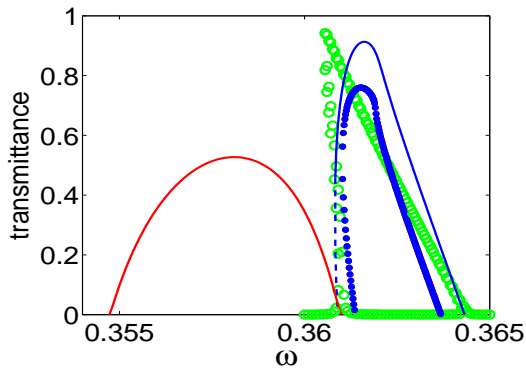}
\caption{(Color online) Transmittances vs frequency for $E_L=E_R=0.035$.
The CMT based transmittance with degenerated dipole modes is shown by solid lines where
red line shows transmission from the right to the left, and blue line shows transmission from
the left to the right. The case
$\omega_1a/2\pi c=0.3658, \omega_2a/2\pi c=0.3650$ is presented by blue closed circles.
Green open circles shows transmittance in PhC structure shown in Fig. \ref{fig1} with $P=0.7W/a$. }
\label{fig5}
\end{figure}

In Fig. \ref{fig5} we present transmittance of the system versus frequency
for the bifurcated solution of the CMT
equations (\ref{A1A2dip}) and compare the results with numerical calculation of
the Maxwell equations for light transmission through the PhC design
shown in Fig. \ref{fig1}. The numerical approach is described in detail in Ref. \cite{BS}.
Patterns of stationary scattering wave function
(real parts of electric field for the TM mode) are presented in Fig. \ref{fig1}
for the frequency $\omega=0.3618$ and the injected power of light $ P=0.7W/a$. This point
belongs to the blue domain in Fig. \ref{fig3} where we have AOD from the left to the right.
Respectively, Fig. \ref{fig1} (a) shows
that when light incidents from
the left the bifurcated solution occurs with both dipole modes excited  to permit the light transmission
through the waveguide.
Fig. \ref{fig1} (b) shows that when light incidents from the right
only single dipole mode is excited similar to the linear case. And therefore
the light flow from the right to the left is blocked.

However numerics in PhC shown in Fig. \ref{fig5} by open green circles does
not reveal the domain of AOD
from the right to the left
in the frequency domain shown in Fig. \ref{fig3} by red or by
solid red line in Fig. \ref{fig5} which follow from the CMT presented above.
The reason is that the approximation of degeneracy of the dipole modes used in the
CMT Eqs. (\ref{A1A2dip}).
When the degeneracy was lifted the transmittance from the right to
the left disappeared as shown in Fig. \ref{fig5} by blue closed circles.
Moreover that gives rise to fine splitting of the transmission peak.

This  work  was  partially  supported  by  RFBR grant 13-02-00497.

\end{document}